\documentclass[12pt,preprint]{aastex}
\usepackage{psfig}

\shorttitle{The Radio Afterglow from the SGR\,1806-20 Giant Flare}
\shortauthors{Taylor et al.}
\begin{document}

\def\HI {H\kern0.1em{\sc i}}
\def\sgr{SGR\,1806-20}
\def\simlt{\mathrel{\hbox{\rlap{\hbox{\lower4pt\hbox{$\sim$}}}\hbox{$<$}}}}
\def\simgt{\mathrel{\hbox{\rlap{\hbox{\lower4pt\hbox{$\sim$}}}\hbox{$>$}}}}

\title{The Growth, Polarization, and Motion
of the Radio Afterglow from the Giant Flare from SGR\,1806-20}

\def\kipac{1} \def\vlba{2} \def\cfa{3} \def\marshall{4}
\def\southampton{5} \def\jive{8} \def\ias{6} \def\bgu{7}
\def\ua{9}

\author{G. B. Taylor \altaffilmark{\kipac,\vlba}, 
J.D. Gelfand \altaffilmark{\cfa}, 
B.M. Gaensler \altaffilmark{\cfa}, 
J. Granot \altaffilmark{\kipac}, 
C. Kouveliotou \altaffilmark{\marshall}, 
R. P. Fender \altaffilmark{\southampton},
E. Ramirez-Ruiz \altaffilmark{\ias}, 
D. Eichler \altaffilmark{\bgu}, 
Y. E. Lyubarsky \altaffilmark{\bgu}, 
M. Garrett \altaffilmark{\jive}, \&
R. A. M. J. Wijers \altaffilmark{\ua}
}

\altaffiltext{\kipac}{Kavli Institute of Particle Astrophysics and Cosmology,
Menlo Park, CA 94025, USA}
\altaffiltext{\vlba}{National Radio Astronomy Observatory, Socorro, NM 87801, USA}
\altaffiltext{\cfa}{Harvard-Smithsonian Center for Astrophysics, 60 
Garden Streen, Cambridge, MA 02138, USA}
\altaffiltext{\marshall}{NASA/MSFC, XD-12, NSSTC, 320 Sparkman Dr., Huntsville, AL
35805, USA}
\altaffiltext{\southampton}{School of Physics and Astronomy, University 
of Southampton, Highfield, Southampton SO17 1BJ, UK}
\altaffiltext{\ias}{IAS, Einstein Drive, Princeton, NJ 08540, USA}
\altaffiltext{\bgu}{Dept. of Phys., BGU, P.O. Box 653, Be\'er Sheva 84105, Israel}
\altaffiltext{\jive}{Joint Institute for VLBI in Europe, Postbus 2, 
7990 AA, Dwingeloo, The Netherlands}
\altaffiltext{\ua}{Astronomical Institute ``Anton Pannekoek'', University
of Amsterdam, Kruislaan 403, 1098 SJ, Amsterdam, The Netherlands}

\begin{abstract}
  
  The extraordinary giant flare (GF) of 2004 December 27 from the soft
  gamma repeater (SGR) 1806-20 was followed by a bright radio
  afterglow.  We present an analysis of VLA observations of this radio
  afterglow from \sgr, consisting of previously reported 8.5 GHz data
  covering days 7 to 20 after the GF, plus new observations at 8.5 and
  22 GHz from day 24 to 81.  We detect motion in the flux
  centroid of the afterglow, at an average velocity of 0.26 $\pm$ 0.03
  c (assuming a distance of 15 kpc) at a position angle of
  $-45^\circ$.  This motion, in combination with the growth and
  polarization measurements, suggests an asymmetric outflow,
  mainly from one side of the magnetar. We find a
  deceleration in the expansion, from $\sim$9 mas/day to $<$5
  mas/day.  The time of deceleration is roughly coincident with the
  rebrightening in the radio light curve, as expected to result when
  the ejecta from the GF sweeps up enough of the external medium, and
  transitions from a coasting phase to the Sedov-Taylor regime.  The
  radio afterglow is elongated and maintains a 2:1 axis ratio with an
  average position angle of $-$40$^\circ$ (north through east),
  oriented perpendicular to the average intrinsic linear polarization
  angle.  
\end{abstract}

\keywords{pulsars: individual (SGR 1806-20) -- stars: neutron -- stars:flare
-- stars: winds,outflows -- radio continuum: general}

%\vfill\eject
\section{Introduction}

The spectacular giant flare (GF) of 2004 Dec. 27 from the soft gamma
repeater (SGR) 1806-20 is believed to have originated from a violent
magnetic reconnection event in this magnetar \citep{pal05,hur05}.
This sudden energy release of more than $10^{46}\;$ergs in gamma-rays
(assuming isotropic emission at a distance of 15 kpc - Corbel \& Eikenberry 2004; McClure-Griffiths \& Gaensler 2005\nocite{CE04,MCC05}) managed to eject
a significant amount of baryons, probably accompanied by some pairs
and magnetic fields, from the neutron star
\citep{pal05,gel05,gra05}. As this outflow interacted with the
external medium, it powered an expanding radio afterglow
\citep{cam05,gae05} at least 500 times more luminous than the only
other radio afterglow detected from an SGR GF \citep{fra99}.  After a
steep decay \citep[$\sim t^{-2.7}$;][]{gae05}, a
rebrightening in the radio light curve was seen, starting at $t\sim
25\;$days and peaking at $t\sim 33\;$days \citep{gel05}, followed
by a shallower decay. This is most naturally explained by the
transition from free expansion to the Sedov-Taylor phase, which occurs
when a sufficient mass of ambient medium is swept up
\citep{gel05,gra05}.

In this {\it Letter} we report on $8.5\;$GHz and $22\;$GHz radio
observations with the Very Large Array (VLA) of the NRAO\footnote{The
National Radio Astronomy Observatory is operated by Associated
Universities, Inc., under cooperative agreement with the National
Science Foundation.} between 7 and 81 days after the GF.  These
observations are used to measure the size, shape, motion and
polarization properties of the radio afterglow. 

%Assuming that \sgr\ is
%at a distance of $15d_{15}\;$kpc, it is useful to
%remember that $1\;$mas corresponds to $15d_{15}\;$AU or $2.25\times
%10^{14}d_{15}\;$cm. The radio observations are described in \S
%\ref{sec:obs}. In \S \ref{sec:fit} we perform fits for the shape and
%size of the radio image using the visibility data. Our results are
%presented in \S \ref{sec:res} and discussed in \S \ref{sec:dis}.

\section{Observations}\label{sec:obs}

VLA observations of \sgr\ began $6.9\;$days after the GF with the VLA
in its A configuration.  Here we report all $8.5\;$GHz and $22\;$GHz
observations up through day 81 (see Table 1).  The first 20 days of
monitoring with a host of radio telescopes including the VLA have
previously been described by \cite{gae05} and by \cite{cam05b}.
Absolute flux calibration was obtained from a short observation of
3C286 during each run. Phase calibration was determined by
observations of the strong (0.75 Jy) but somewhat distant (5.78 deg)
calibrator PMN J1820-2528, or (from 2005 Jan. 16 on) the nearby (0.77
deg), and moderately strong (0.32 Jy) calibrator TXS J1811-2055 with a
cycle time of 3.5 minutes.  From Jan. 16 onwards, the validity of the
phase transfer at 8.5 GHz was checked by short observations of
J1820-2528 every 15 minutes.  In general the coherence was found to be
better than 95\% on J1820-2528.  For all observations except 2005
Jan. 3 the strong and unpolarized source, OQ208, was observed for 1
minute in order to permit solving for the instrumental polarization.
For the data on Jan. 3, leakage terms were transfered from
observations of BL Lac on 2005 Jan. 2.  The absolute polarization
angle was referenced to 3C286 for all epochs.

\section{Model fitting and Error Analysis}\label{sec:fit}

In all observations reported here the radio afterglow of \sgr\ is
smaller than the naturally weighted synthesised beam.  Since the
signal to noise is high, however, it is quite feasible to extract
information about the size and shape of the source by fitting models
to the visibility data.  For each of the epochs we fit a two component
model to the data for the \sgr\ field.  One elliptical,
two-dimensional Gaussian component (with the 6 free parameters given in 
Table 2) describes \sgr\, while a point source (not listed) was
used to describe the radio nebula associated with the LBV star
approximately 14 arcseconds to the East \citep{fra97}.  Other models
for the radio afterglow, including an elliptical ring, a uniform
sphere, an elliptical disk, and two point sources, were tried but not
found to provide a better fit.  Fitting the VLA data to
an elliptical ring, or disk at any epoch increases the derived size by
a factor $\sim$1.16 and $\sim$1.66 respectively as expected
\citep{pea99}.  The model fitting was performed in both MIRIAD (task
UVFIT) and
Difmap and found to agree to within the uncertainties.  We adopt the
MIRIAD fits (Table 2) and the estimated statistical errors.
As in
\cite{tay04} the error of the size was checked with Monte-Carlo
simulations of the data using identical ($u$,$v$) coverage, similar
noise properties, and a Gaussian component of known size added.  The
simulations confirm the error estimates quoted by MIRIAD, and agree
with errors estimated from the signal to noise ratio and the
synthesised beam shape.

In the early epochs there is some evidence from the MERLIN and VLBA
observations \citep{fen05} that the morphology of the source is more
complicated than an elliptical Gaussian, and may experience rapid
changes in the location of the peak emission.  These changes in the
suface brightness could cause shifts in the centroid of our model
fits, and deviations in the fitted size.  For this reason we have
added a 10\% error in quadrature to the measured size of all points,
though the error may be larger in the earlier measurements.  At later
times 1.4 GHz MERLIN observations seem to be more consistent with a smooth
elliptical Gaussian.

\section{Results}\label{sec:res}

\subsection{Polarization}\label{sec:polar}

Linear polarization from the radio afterglow was detected
during the first 20 days after the GF at 8.5 GHz \citep{gae05}.  Thereafter
we were only able to measure upper limits on the polarization (Fig.~1).
The polarization is found to be 2.1\% on day 7 and to decrease
to a minimum of 1.1\% on day 10.  At that time the linear polarization
began to increase steadily to a maximum value of 3.4 \% while the polarization
angle swung rapidly from 4$^\circ$ to 40$^\circ$.  The polarization
falls below our detection limit of 2\% around the time of the 
rebrightening in the light curve.  Limits as late
as 55 days after the GF are below 2\%.

\subsection{Expansion}\label{sec:exp}

In Fig.~2 we plot the geometric mean diameter of the elliptical Gaussian
model fits.  These fits show the expansion of the radio afterglow from
\sgr\ over the first 81 days after the GF.  As reported in
\cite{gae05} \sgr\ was clearly resolved in the earliest VLA
observations taken 7 days after the GF with a diameter of $\sim$57 mas
(mean full-width).  MERLIN observations \citep{fen05} reveal that the
source could be asymmetric.  The size and position angle of the MERLIN
extension at $-$31 degrees is roughly consistent with our average
value of $-$40 $\pm$ 20.  There is some possible evidence for a
gradual swing in the position angle of the VLA data (Table 2), and we
will investigate its significance in a future paper.  Assuming a
one-sided expansion (as argued in \S 5), the apparent velocity
required to reach a size of 57 mas in 6.9 days is 8.3 $\pm$ 0.9 mas/day
(0.72 $\pm$ 0.10 $d_{15}c$).  

After 30 days \citep[the time of the rebrightening reported
by][]{gel05} the radio afterglow had grown to $\sim260\;$mas
(full-width of the geometric mean of the major and minor axes).
Between 7 and 30 days the growth of the radio nebula from 57 mas to
260 mas corresponds to an average expansion velocity of 9.0 $\pm$ 1.6
mas/day (0.78 $\pm$ 0.14 $d_{15}c$).  After this time, the growth
appears to slow so that the average velocity between day 30 and the
last day of observations reported here is 1.0 $\pm$ 2 mas/day ($<0.4
d_{15}c$) where the source size reaches $\sim$322 mas.  This expansion
is in agreement with the MERLIN size estimate of $\sim$220 mas (mean
full-width), 56 days after the GF.

Following \cite{gel05} (see their Eq. 4), we fit to the data from day
9 onwards a supersonically expanding spherical shell that is
decelerated as it sweeps up material.  
While the deceleration of an anisotropic outflow
might be somewhat different than in the spherical case, the latter may    
still serve as a rough approximation.
The fit (reduced $\chi^2$ of 0.76; shown as the solid line in Fig.~2)
implies a deceleration time of 40 $\pm$ 13 days after the GF,
consistent with the time of the peak rebrightening at day $\sim$30 and
the deceleration time of $\sim$46 days derived from the rebrightening
\citep{gel05}.  We also fit a constant expansion (5.6 $\pm$ 0.6
mas/day) to the data and obtain a reduced $\chi^2$ of 1.22.  An F-test
gives a probability of 2\% that the constant expansion is an equally
valid description of the data.  A broken power-law actually fits much
better than either model (reduced $\chi^2$ = 0.06), but requires both
acceleration and deceleration of the explosion.

%The low $\chi^2$
%value of the broken power-law may indicate that we have been too
%conservative with our error estimation.

\subsection{Proper Motion}\label{sec:motion}

Good astrometry was obtained for the radio afterglow from \sgr\ on all
days of the observations except 2005 Jan. 3, Jan. 10, and Feb. 7 via
phase referencing to a nearby calibrator.  A
combination of a long cycle time (15 min), distant calibrator
(J1820-2528), and poor atmospheric phase stability resulted in a large
systematic position error on Jan. 3, though changes in the relative
brightness of different parts of the image \citep{fen05} may also have
affected the centroid position.  On Jan. 10 the low elevation of the
observations forced us to employ a distant calibrator with a poor
position.  On Feb. 7 poor weather caused unstable phase conditions
such that the coherence estimated on J1820-2528 was only 36\%.

The centroid of the radio afterglow from \sgr\ is found to shift by
$\sim$200 mas over the course of 70 days of observations (Fig.~3).  We
have decomposed this shift into x and y components (Table~2) and
performed least squares fits to the motion.  The radial proper motion
is 3.0 $\pm$ 0.34 mas/day at a position angle of $-$44 $\pm$ 6
$^\circ$ (measured north through east).  This motion corresponds to
$0.26 \pm 0.03 d_{15}c$.  There is some indication that the time of
fastest proper motion also corresponds to the time of fastest growth.

\section{Discussion}\label{sec:dis}

The motion of the radio flux centroid is along the major axis of the
source and is roughly half of the growth rate.  This may be naturally
explained by a predominantly one-sided outflow, which produces a radio
nebula extending from the location of the magnetar out to an
increasing distance in the direction of the ejection.  This suggests
that either the catastrophic reconfiguration of the magnetic field
which caused the GF was relatively localized, rather than a global
event involving the whole magnetar \citep[c.f.,][]{eic02}, or that the
baryonic content of the ejecta is asymmetric.  It is also possible
that the environment plays a role in collimating the outflow.

\begin{figure}
%%\epsscale{0.2}
%%\plotone{finf.eps}
%\special{psfile=f1.eps hoffset=-10 voffset=-50 hscale=80.0 vscale=80.0}
\vspace{-1.0cm}
\plotone{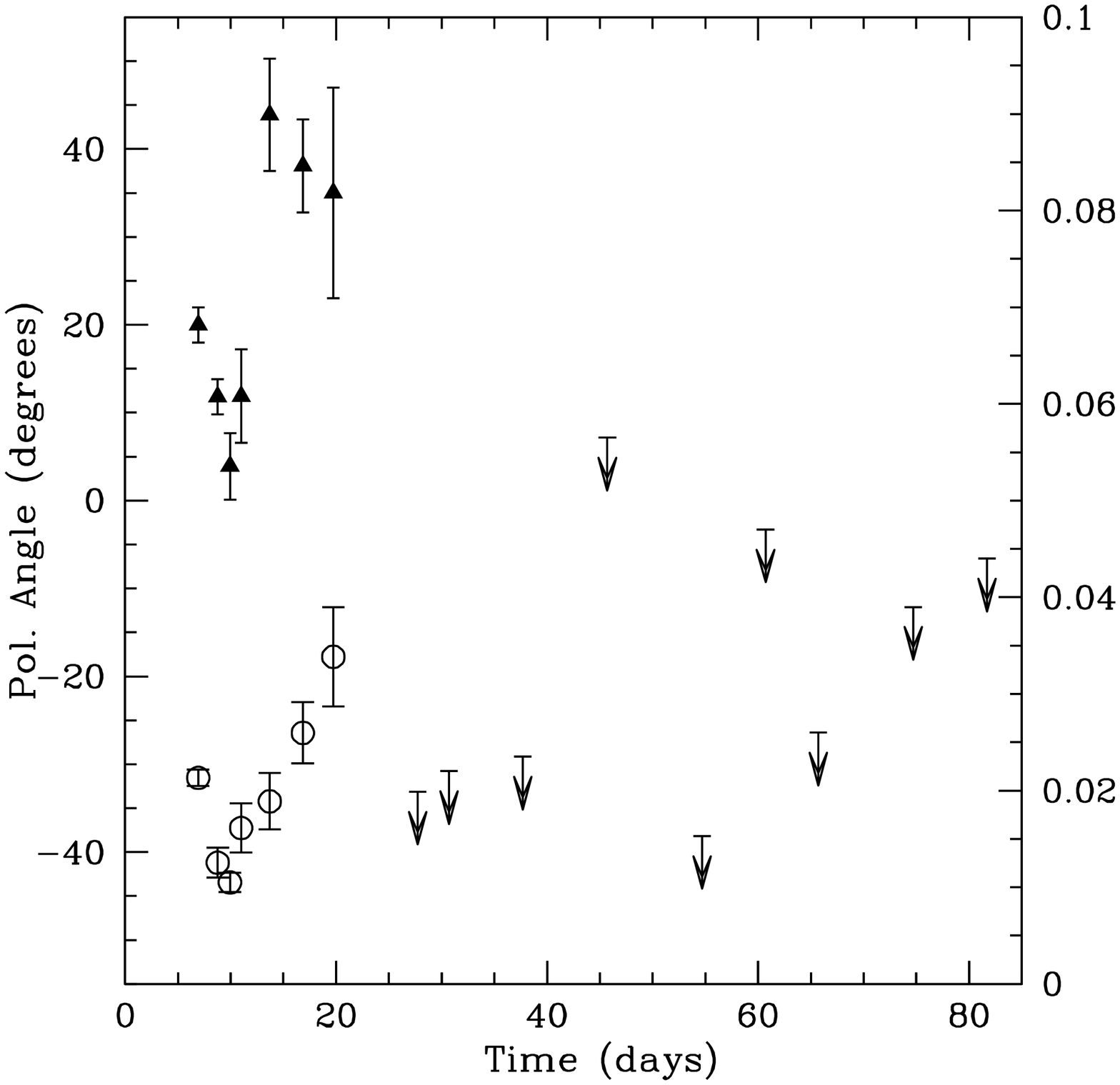}
\vspace{-1.8cm}
\caption{Linear fractional polarization (circles; right y-axis) and polarization 
angles (triangles; left y-axis) for the radio afterglow from \sgr\ as a function of time at 8.5 GHz.
All polarization angles have been corrected for the observed RM of 
272 $\pm$ 10 rad m$^{-2}$ \citep{gae05}.  Limits on fractional 
polarization are drawn at 3$\sigma$.}
\label{fig:pol}
\end{figure}

The position angle of the linear polarization is roughly perpendicular
to the major axis of the image and to the direction of motion of the
flux centroid. This naturally arises for a shock-produced magnetic
field, which is tangled predominantly within the plane of the shock
\citep{ml99}, because of the elongated shape of the emitting region
and due to projection effects \citep{gae05}. Alternatively, this might
be caused by shearing motion along the sides of the one-sided outflow,
which can stretch the magnetic field in the emitting region along its
direction of motion.  The degree of polarization decreased at about
the same time as the deceleration and rebrightening in the light curve
(see Fig. \ref{fig:pol}). As the rebrightening is attributed to the
emission from the shocked external medium becoming dominant
\citep{gel05,gra05}, this suggests a lower polarization of this
emission component. This, in turn, suggests that the magnetic field in
the shocked ISM is less ordered than in the shocked ejecta and/or
shocked external shell.

In the first 30 days, the outer edge of the radio afterglow moves away
from the magnetar position at an apparent velocity of 0.8 c.  The
intrinsic velocity could be lower depending on the unknown inclination
of the outflow. The minimum velocity is 0.62 c for an angle of the
outflow with the line-of-sight of 51$^\circ$.  This is in agreement
with the high escape velocity of 0.5 c for a neutron star.  At these
trans-relativistic velocities (Lorentz factor 1.3) there is a modest
increase in the total kinetic energy of the outflow.  Compared to
previous estimates based on isotropic outflows \citep{gel05}, the
energy is increased by a factor of 2-3 owing to the factor 2 higher
velocity at the outer edge, but lower velocities elsewhere
\citep{gra05}.  Combining these two factors leads to a revised
estimate for the total kinetic energy in the ejecta of $\sim$10$^{45}$
ergs.  By momentum conservation, a one-sided outflow of $10^{24.5}$ g
\citep{gra05} at 0.62 c imparts a kick to the magnetar of 30 cm s$^{-1}$.

\section{Conclusions}

We report a deceleration in the observed expansion of the radio
afterglow produced by the 2004 Dec. 27 Giant Flare from \sgr.  We also
find a proper motion for the radio afterglow roughly aligned with its
major axis and perpendicular to the average polarization angle.  These
observations support the idea of an asymmetric explosion on one side
of the magnetar.  The polarization data place significant constraints
on the magnetic field structure in the shocked ejecta and ISM.
Measurements with the VLA continue, and will be presented in a future
paper.

\acknowledgments

%GBT thanks the Kavli Institute for Particle Astrophysics and
%Cosmology for hospitality and support. 

We thank an anonymous referee for constructive comments.  J.D.G. and B.M.G.
acknowledge the support of NASA through LTSA grant NAG5-13032.
J.G. acknowledges support from the US Department of Energy 
under contract number DE-AC03-76SF00515.
D.E. acknowledges support from the Israel-U.S. BSF, the ISF,
and the Arnow Chair of Physics. Y.L acknowledges support from the
German-Israeli Foundation. E.R.R is sponsored by NASA through a
Chandra Postdoctoral Fellowship award PF3-40028. R.A.M.J.W
acknowledges support from NWO.

\begin{figure}
%\vspace{20.0cm}
%%\epsscale{0.2}
%%\plotone{sizefit.eps}
%%\special{psfile=f2.eps angle=20 hoffset=-10 voffset=-50 hscale=50.0 vscale=50.0}
\vspace{-0.5cm}
\plotone{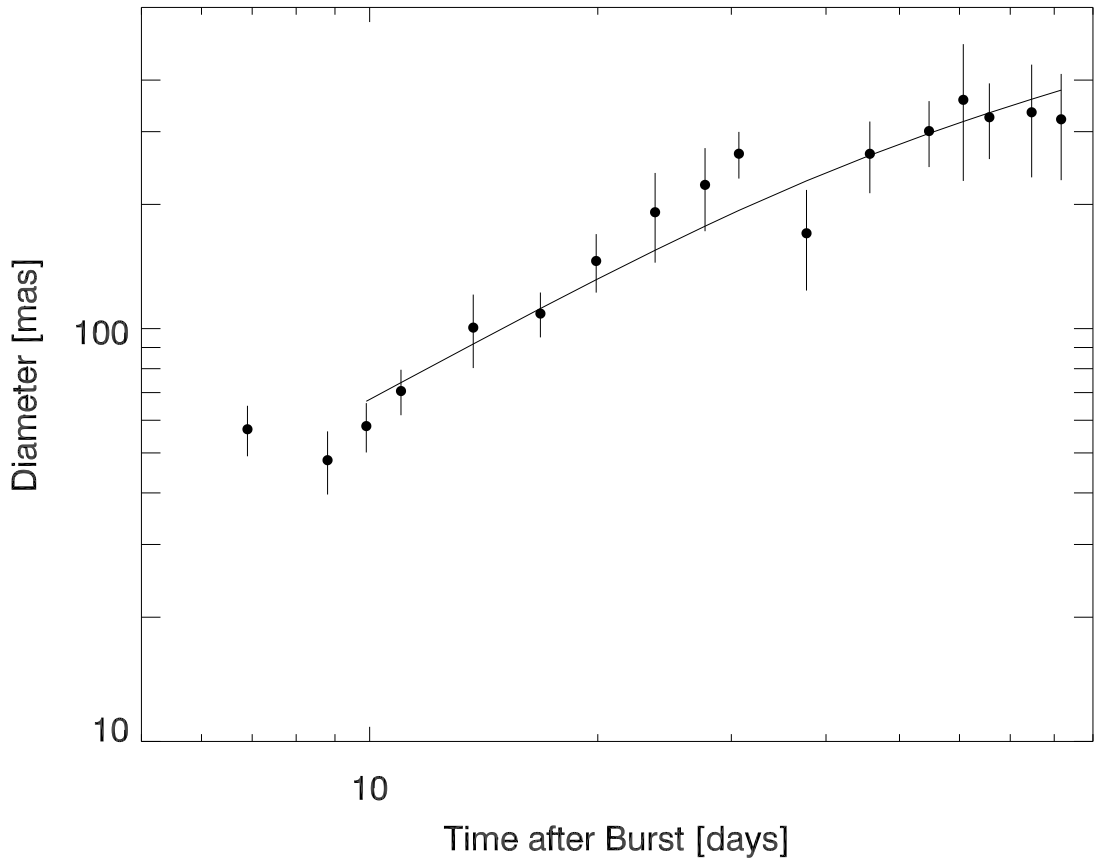}
\vspace{-4.0cm}
\caption{Expansion of the radio afterglow from \sgr\ as a function of time.
The size shown is the geometric mean of the major and minor axes of 
the best fitting elliptical Gaussian for each observation. The solid
line is a fit of a supersonically expanding shell model as described
by Eq.~4 of \cite{gel05}. This model does not take into account the
collimation and proper motion of the source, but provided that these
are not extreme, it illustrates the deceleration due to mass loading by the
external medium. }
\label{fig:sizemod}
\end{figure}
\clearpage

\begin{figure}
%\vspace{20.0cm}
%\epsscale{0.7}
\vspace{-1.0cm}
\plotone{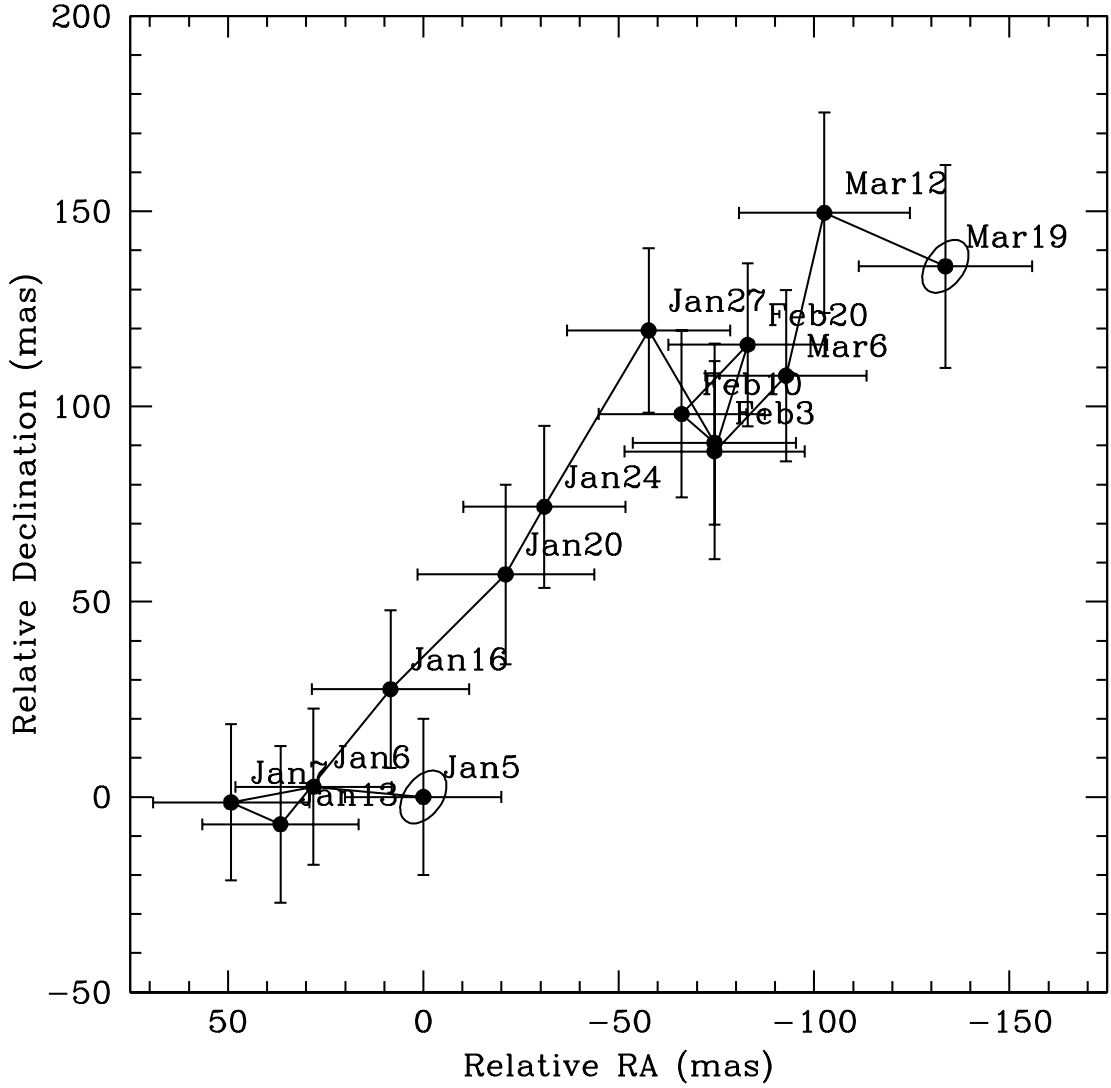}
\vspace{-1.8cm}
%\special{psfile=skyplot.eps hoffset=-10 voffset=-50 hscale=80.0 vscale=80.0}
\caption{The trajectory of the afterglow of \sgr.  Dates are labeled.
The small ellipses denote the first and last days used.}
\label{FIG3}
\end{figure}
\clearpage

\begin{deluxetable}{lrrrrrrrrr}
\tabletypesize{\footnotesize}
\tablecolumns{10}
\tablewidth{0pt}
\tablecaption{Observational Summary\label{Observations}}
\tablehead{\colhead{Date}&\colhead{$t$}&\colhead{Freq.}&
\colhead{Phase}&
\colhead{Time}&\colhead{RMS noise}&\colhead{Array}&\colhead{$B_{\rm min}$}&
\colhead{$B_{\rm maj}$}&\colhead{$B_{\rm P.A.}$}\\
\colhead{}&\colhead{(days)}&\colhead{(GHz)}&\colhead{Calibrator}&
\colhead{(min)}&\colhead{($\mu$Jy/beam)}&\colhead{Config.}&\colhead{(mas)}&\colhead{(mas)}&
\colhead{(deg.)}}
\startdata
2005 Jan 3 & 6.9 & 8.5 & J1820{\tt -}2528 & 12 & 60 & A & 222 & 458 & 16 \\
2005 Jan 5 & 8.8 & 8.5 & J1820{\tt -}2528 & 15 & 70 & A & 213 & 542 & $-$27 \\
2005 Jan 6 & 9.9 & 8.5 & J1820{\tt -}2528 & 34 & 34 & A & 233 & 430 & 19 \\
2005 Jan 7 & 11.0 & 8.5 & J1820{\tt -}2528 & 18 & 61 & A & 233 & 717 & 41 \\
2005 Jan 10 & 13.7 & 8.5 & J1751{\tt -}2524 & 26 & 45 & A & 228 & 811 & $-$41 \\
2005 Jan 13 & 16.8 & 8.5 & J1820{\tt -}2528 & 38 & 25 & A & 295 & 597 & 29\\
2005 Jan 16 & 19.9 & 8.5 & J1811{\tt -}2055 & 28 & 31 & A & 408 & 605 & $-$42\\
2005 Jan 20 & 23.8 & 22.5 & J1811{\tt -}2055 & 37 & 59 & BnA & 190 & 317 & $-$75\\
2005 Jan 24 & 27.7 & 8.5 & J1811{\tt -}2055 & 21 & 30 & BnA & 451 & 1004 & $-$60\\
2005 Jan 27 & 30.7 & 8.5 & J1811{\tt -}2055 & 32 & 36 & BnA & 382 & 1346 & 53\\
2005 Feb 3 & 37.7 & 8.5 & J1811{\tt -}2055 & 21 & 40 & BnA & 437 & 1062 & $-$60\\
2005 Feb 7 & 41.7 & 8.5 & J1811{\tt -}2055 & 21 & -- & BnA & -- & -- & -- \\ 
2005 Feb 11 & 45.7 & 8.5 & J1811{\tt -}2055 & 17 & 40 & BnA & 401 & 1328 & $-$55\\
2005 Feb 20 & 54.7 & 8.5 & J1811{\tt -}2055 & 111 & 15 & B & 736 & 1323 & $-$14\\
2005 Feb 26 & 60.7 & 8.5 & J1811{\tt -}2055 & 17 & 48 & B & 706 & 1465 & $-$22\\
2005 Mar 4 & 66.7 & 8.5 & J1811{\tt -}2055 & 66 & 36 & B & 720 & 1296 & 3 \\
2005 Mar 12 & 74.7 & 8.5 & J1811{\tt -}2055 & 26 & 42 & B & 730 & 1305 & 3 \\
2005 Mar 19 & 81.7 & 8.5 & J1811{\tt -}2055 & 37 & 44 & B & 736 & 1351 & 14 \\
\enddata 
\tablenotetext{*}{NOTE - VLA data from Feb. 7 was unusable
due to poor observing conditions.  Feb. 11 includes data \\ taken on
Feb 10 and Feb 12.  Feb. 20 includes data taken on Feb. 19 and Feb.
21. Column 2 gives $t$, the time after the GF, column 5 refers to 
the total integration time on source, and $B_{\rm maj}$, $B_{\rm min}$, and $B_{\rm P.A.}$ describe the naturally weighted 
synthesized restoring beam measured north through east. }
\end{deluxetable}

\bigskip

\begin{deluxetable}{lrrrrrrrr}
\tabletypesize{\footnotesize}
\tablecolumns{9}
\tablewidth{0pt}
%\tabcolsep{0.1in}
\tablecaption{Model Fitting and Polarimetry Results\label{Fits}}
\tablehead{
\colhead{$t$} & \colhead{Flux} & \colhead{$\Delta$ x} & 
\colhead{$\Delta$ y} & \colhead{$\theta_{\rm M}$} & \colhead{Axial} & 
\colhead{$\theta_{\rm PA}$} & 
\colhead{Pol.} & \colhead{$\phi$} \\
\colhead{(days)} & \colhead{(mJy)} & \colhead{(mas)} & \colhead{(mas)} &
\colhead{(mas)} & \colhead{Ratio} & \colhead{(deg.)} & \colhead{(\%)} & 
\colhead{(deg.)}}
\startdata
% 6.9 & 54.70 & $-$11  &  96 &  80 & 2.1 $\pm$ 0.1 & 20 $\pm$ 2 \\
6.9 & 54.59$\pm$0.09 & --  & -- & 79.4$\pm$0.9 & 0.52$\pm$0.06 &  $-$58$\pm$2 & 2.1 $\pm$ 0.1 & 20 $\pm$ 2 \\
8.8 & 32.30$\pm$0.09 & 0   &  0 & 67.8$\pm$4.9 & 0.50$\pm$0.12 &  $-$65$\pm$8 & 1.3 $\pm$ 0.2 & 12 $\pm$ 2  \\
9.9 & 23.68$\pm$0.04 & 28 &  2  & 78.6$\pm$1.2 & 0.55$\pm$0.06 &  $-$52$\pm$3 & 1.1 $\pm$ 0.1  & 4 $\pm$ 4 \\
11.0 & 16.78$\pm$0.06 & 47 &$-$5 & 85.8$\pm$2.4 & 0.68$\pm$0.03 & $-$68$\pm$12 & 1.6 $\pm$ 0.3 & 12 $\pm$ 5 \\
13.7 & 9.75$\pm$0.05 & -- &  -- & 120.3$\pm$16.3 & 0.70$\pm$0.10 & $-$59$\pm$12 & 1.9 $\pm$ 0.3 & 44 $\pm$ 6 \\
16.8 & 5.65$\pm$0.04 & 37 &$-$9 & 121.6$\pm$5.9  & 0.80$\pm$0.13 & $-$87$\pm$24 & 2.6 $\pm$ 0.3 & 38 $\pm$ 5 \\
19.9 & 4.18$\pm$0.05 & 10 &  26 & 197.7$\pm$12.8 & 0.54$\pm$0.08 & $-$54$\pm$8 & 3.4 $\pm$ 0.5 & 35 $\pm$ 12 \\
23.8 & 1.62$\pm$0.12 & $-$19 & 56 & 208.9$\pm$40.7 & 0.84$\pm$0.25 & $-$31$\pm$46 &     --        &  -- \\
27.7 & 3.24$\pm$0.06 & $-$36 & 71 & 276.1$\pm$42.3 & 0.65$\pm$0.12 & $-$44$\pm$11 & $<$2.0 &  \\
30.7 & 3.93$\pm$0.06 & $-$68 & 111 & 292.3$\pm$17.7 & 0.82$\pm$0.22 & $-$26$\pm$31 & $<$2.2 &  \\
37.7 & 3.22$\pm$0.06 & $-$70 & 91 & 258.3$\pm$37.5 & 0.43$\pm$0.26 & $-$15$\pm$16 & $<$2.4 &  \\
45.7 & 2.60$\pm$0.05 & $-$61 & 97 & 346.4$\pm$38.9 & 0.59$\pm$0.11 & $-$26$\pm$10 & $<$5.7 &   \\
54.7 & 2.03$\pm$0.03 & $-$78 & 109 & 352.5$\pm$41.7 & 0.73$\pm$0.10 & $-$21$\pm$11 & $<$1.5 & \\
60.7 & 1.78$\pm$0.07 & $-$67 & 75 & 461.4$\pm$117 & 0.60$\pm$0.30 & $-$28$\pm$17 & $<$4.7 &  \\
65.7 & 1.72$\pm$0.04 & $-$92 & 107 & 446.8$\pm$50.6 & 0.53$\pm$0.09 & $-$12$\pm$7 & $<$2.6 &   \\
74.7 & 1.55$\pm$0.06 & $-$113 & 128 & 446.1$\pm$90.3 & 0.56$\pm$0.16 & $-$11$\pm$13 & $<$3.9 &   \\
81.7 & 1.39$\pm$0.05 & $-$135 & 141 & 459.1$\pm$78.6 & 0.49$\pm$0.22 & $-$23$\pm$15 & $<$4.4 &   \\
\enddata 
\tablenotetext{*}{NOTE - Positions are relative to that derived
on Jan. 5 which is RA 18 08 39.3418, DEC $-$20 24 39.827 (J2000).  
The positions of Jan. 3 and 10 are excluded for reasons
described in the text.  The errors quoted on the flux densities are
only statistical, and will be discussed in a future paper.   
Position errors are dominated by a $\sim$20 mas
systematic uncertainty in the astrometry. The source size 
is described by the major axis, $\theta_{\rm M}$, axial ratio,
and position angle $\theta_{\rm P.A.}$. Note that
the PA at 13.7 days after the burst differs substantially 
from the value of 62 $\pm$ 14
deg shown in Fig 3 of Gaensler et al (2005). The new fit presented here
seems more consistent with the PA's seen at adjacent epochs. A subsequent
paper will fully investigate this possible discrepancy.
The source
polarization is described by the fractional polarization in \% and
the electric vector polarization angle, $\phi$, where $\phi$ has been 
corrected for the observed RM of 272 $\pm$ 10 rad m$^{-2}$ \citep{gae05}}
\end{deluxetable}

\end{document}